\documentclass[aps,prl,twocolumn,showpacs,superscriptaddress]{revtex4-1}
\usepackage{amsmath, amssymb, graphicx, bm, subcaption, mathtools}

\usepackage[usenames,dvipsnames,svgnames,table]{xcolor}
\usepackage[colorlinks=true,linkcolor=RoyalBlue,citecolor=RoyalBlue]{hyperref}

\begin{document}

\title{Magnetic domain wall Skyrmions}
\author{Ran Cheng}
\affiliation{Department of Physics, Carnegie Mellon University, Pittsburgh, PA 15213, USA}
\affiliation{Department of Electrical \& Computer Engineering, University of California, Riverside, CA 92521, USA}

\author{Maxwell Li}
\affiliation{Department of Materials Science \& Engineering, Carnegie Mellon University, Pittsburgh, PA 15213, USA}

\author{Arjun Sapkota}
\affiliation{Department of Physics and Astronomy/MINT Center, The University of Alabama, Tuscaloosa, AL 35487, USA}

\author{Anish Rai}
\affiliation{Department of Physics and Astronomy/MINT Center, The University of Alabama, Tuscaloosa, AL 35487, USA}

\author{Ashok Pokhrel}
\affiliation{Department of Physics and Astronomy/MINT Center, The University of Alabama, Tuscaloosa, AL 35487, USA}

\author{Tim Mewes}
\affiliation{Department of Physics and Astronomy/MINT Center, The University of Alabama, Tuscaloosa, AL 35487, USA}

\author{Claudia Mewes}
\affiliation{Department of Physics and Astronomy/MINT Center, The University of Alabama, Tuscaloosa, AL 35487, USA}

\author{Di Xiao}
\affiliation{Department of Physics, Carnegie Mellon University, Pittsburgh, PA 15213, USA}
\affiliation{Department of Materials Science \& Engineering, Carnegie Mellon University, Pittsburgh, PA 15213, USA}

\author{Marc De Graef}
\affiliation{Department of Materials Science \& Engineering, Carnegie Mellon University, Pittsburgh, PA 15213, USA}

\author{Vincent Sokalski}
\affiliation{Department of Materials Science \& Engineering, Carnegie Mellon University, Pittsburgh, PA 15213, USA}

\begin{abstract}
It is well established that the spin-orbit interaction in heavy metal/ferromagnet heterostructures leads to a significant interfacial Dzyaloshinskii-Moriya Interaction (DMI), which modifies the internal structure of magnetic domain walls (DWs) to favor N\'{e}el over Bloch type configurations.  However, the impact of such a transition on the structure and stability of internal DW defects (e.g., vertical Bloch lines) has not yet been explored.  We present a combination of analytical and micromagnetic calculations to describe a new type of topological excitation called a DW Skyrmion characterized by a $360^\circ$ rotation of the internal magnetization in a Dzyaloshinskii DW.  We further propose a method to identify DW Skyrmions experimentally using Fresnel mode Lorentz TEM; simulated images of DW Skyrmions using this technique are presented based on the micromagnetic results.
\end{abstract}

\maketitle

\textit{Introduction.}---The discovery of a large Dzyaloshinskii-Moriya Interaction (DMI)~\cite{Dzyaloshinsky1958,Moriya1960} in bulk magnetic crystals~\cite{Yu2010,Huang2012} and thin films with structural inversion asymmetry~\cite{Thiaville2012,Hrabec2014,Pellegren2017,Lau2018} has led to a fervent rebirth of research on magnetic bubble domains in the form of smaller particle-like features called Skyrmions, which are minimally defined as having an integer-valued topological charge $Q$, computed from $4\pi\,Q=\int\mathrm{d}x\mathrm{d}y\,\mathbf{m} \cdot \left( \partial_x \mathbf{m} \times \partial_y \mathbf{m} \right)$, where $\mathbf{m}$ is the unit magnetization vector.  Although non-trivial to calculate, there is an inherent energy barrier associated with the annihilation of such an object when $Q$ goes to $0$ --- something widely referred to as topological protection.  The combination of a large DMI, which yields smaller, more stable Skyrmions, and a related spin-orbit coupling phenomenon, viz., the spin Hall effect~\cite{Hirsch1999,Liu2012,Emori2013,Hoffman2013}, makes the prospect of using Skyrmions for energy-efficient memory and computing attractive~\cite{Jiang2015,Woo2016}.

\begin{figure}
	\includegraphics[width = 3.4in]{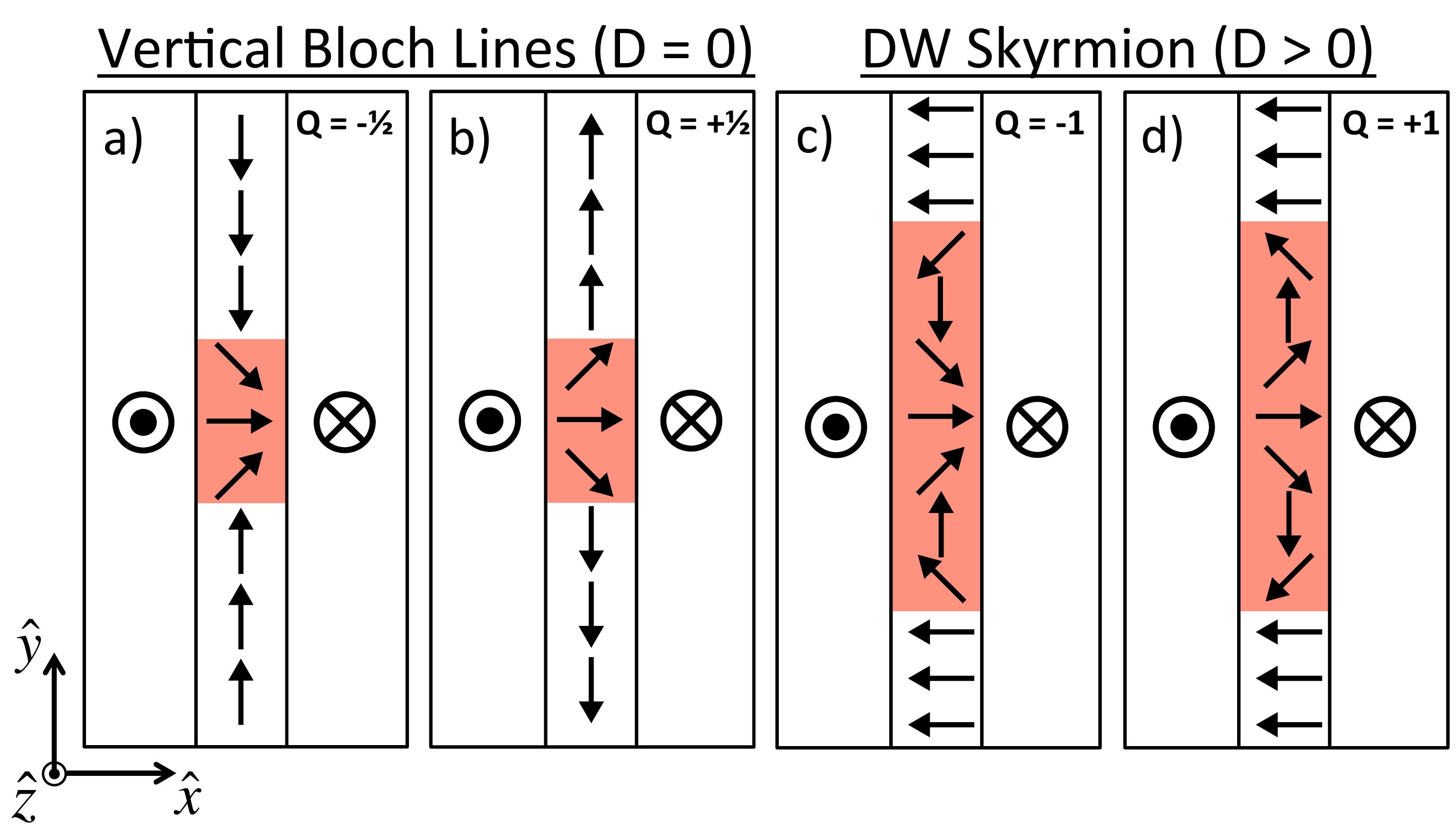}
	\caption{Comparison of vertical Bloch lines (a-b) and DW Skyrmions (c-d), which result from interfacial DMI.} 
	\label{Figure1}
\end{figure}

Here, we present a manifestly different type of topologically protected magnetic excitation called a domain wall (DW) Skyrmion, which has previously been considered under field-theory context~\cite{Jennings2013,Gudnason2014}.  It describes a $360^{\circ}$ wind of the DW's internal magnetization along the wall profile and has a topological charge of $\pm1$. In the absence of DMI, DWs in thin films with perpendicular magnetic anisotropy tend to form the Bloch configuration~\cite{Thiaville1995}.  In these walls, it is common to encounter topological defects ($Q = \pm \frac{1}{2}$) characterized by $180^\circ$ transitions called vertical Bloch lines (VBLs), as shown in Fig.~\ref{Figure1}a-b~\cite{Malozemoff1972,Slonczewski1974}, which were once considered in their own right for computer memory~\cite{Konishi1983,Konishi1984}.  Adding a sufficiently strong interfacial DMI, however, will favor a N\'{e}el-type DW with preferred chirality, known as the Dzyaloshinskii DW~\cite{Thiaville2012}.  Correspondingly, a VBL in the presence of DMI will become a DW Skyrmion as schematically illustrated in Fig.~\ref{Figure1}c-d.  In contrast to conventional Skyrmions that can propagate along any direction in 2D and are subject to the Skyrmion Hall effect~\cite{Jiang2016}, a DW skyrmion can only move in reconfigurable 1D channels defined by the network of magnetic DWs.  Moreover, the interfacial DMI substantially reduces the exchange length along the DW, resulting in a DW Skyrmion that is much smaller than its VBL predecessor --- an observation analogous to 2D Skyrmions and magnetic bubbles.  It is worth noting that unlike the conventional 2D Skyrmions that can form a lattice as the ground state~\cite{Muhlbauer915}, DW Skyrmions can only be metastable excitations. Their existence in systems with a strong DMI has not yet been investigated.

This Letter describes the static properties of DW Skyrmions and proposes a methodology to identify them experimentally.  We begin with an analytical solution of the DW Skyrmion profile obtained by energy minimization, which is found to match well with micromagnetic solutions.  Based on the micromagnetic output, lorentz transmission electron microscopy (LTEM) images are simulated showing that DW Skyrmions should present a clear signature in the Fresnel observation mode.

\textit{Analytical calculations.}---We choose Cartesian coordinates such that the DW normal is along $x$ and the film normal is $z$ (Fig.~\ref{Figure1}). In the thin-film approximation, by assuming that the system is uniform in the thickness direction and infinite along $y$, we have the free energy in the continuum limit as
\begin{align}
 \frac{E}{t_F}=\int &\mathrm{d}x\mathrm{d}y \left\{ A\sum_i|\partial_i\bm{m}|^2 +D \bm{m}\cdot[(\hat{z}\times\bm{\nabla})\times\bm{m}] \right. \notag\\
 &-K m_z^2+\frac{\ln2}{2\pi}\frac{t_F}{\lambda}\mu_0M_s^2[\hat{\bm{n}}(y)\cdot\bm{m}]^2 \bigg\}, \label{eq:energy}
\end{align}
where $t_F$ is the film thickness, $A$ is the exchange stiffness, $i=x,y,z$, $D$ is the DMI, $M_s$ is the saturation magnetization, and $K=K_u-\mu_0M_s^2/2$ is the effective perpendicular magnetic anisotropy with $K_u$ the intrinsic magneto-crystalline anisotropy. The last term represents the demagnetization energy approximated in the thin-film geometry~\cite{Thiaville2012,Tarasenko1998,Slonczewski1974}, where $\lambda=\sqrt{A/K}$ is the exchange length~\cite{Hubert1998} and $\hat{\bm{n}}(y)$ is a \textit{local} normal vector of the DW to account for distortion ~\cite{Sonin1986} in the presence of an internal topological defect.

To solve for $\bm{m}=\bm{m}(x,y)$, we parameterize the magnetization vector in spherical coordinates as
\begin{equation}
\bm{m}=\{ \sin\theta\cos\phi,\ \sin\theta\sin\phi,\ \cos\theta \}.
\end{equation}
In the absence of DMI ($D=0$), minimizing the free energy leads to a standard soliton profile: $\theta=2\arctan\exp(x/\lambda)$ and $\phi=\pm\pi/2$ with $\pm$ representing a Bloch wall of either chirality.

Adding a strong DMI will overcome the demagnetization energy, leading to N\'{e}el walls with either $\phi=0$ or $\pi$. Here we choose $D>0$, thus $\phi=\pi$ at $y=\pm\infty$. As $\phi$ changes, VBLs (Figure~\ref{Figure1}a-b) will gradually transition into DW-Skyrmions (Figure~\ref{Figure1}c-d). Similar to constriction of the DW profile in a VBL due to the increased demagnetization energy~\cite{Sonin1986}, the presence of a DW-Skyrmion also locally deforms the DW profile due to DMI. Because the internal magnetization is inevitably tilted away from the DW normal in a DW Skyrmion, there is a driving force for the DW itself to bend locally as an attempt to recover this energy; this phenomenon is similar to the spontaneous tilting of Dzyaloshinskii DWs identified in~\cite{Boulle2013,Pellegren2017}.  To capture this effect, we adopt a modified Slonczewski ansatz for the profile function that involves two independent variables
\begin{align}
    \theta&=2\arctan\exp\frac{x-q(y)}{\lambda}, \label{eq:theta} \\
    \phi&=\phi(y), \label{eq:phi}
\end{align}
where $\phi(y)$ is the azimuthal angle of the in-plane component of $\bm{m}$ and $q(y)$ denotes the deviation of the DW center from its location in a straight homochiral DW without a VBL or DW Skyrmion; $q(y)$ and $\phi(y)$ are two collective coordinates to be solved by minimizing the total energy. We have neglected a possible $y$-dependence of $\lambda$, which is expected to become significant only for large DMI. Inserting Eqs.~\eqref{eq:theta} and~\eqref{eq:phi} into the energy functional Eq.~\eqref{eq:energy}, noting that the local normal vector $\Hat{\bm{n}}=\{1,q',0\}/\sqrt{1+q'^2}$, and integrating out $x$ from $-\infty$ to $\infty$, we obtain the free energy
\begin{align}
    E=\frac{2t_FA}{\lambda}\int& \mathrm{d}y \left[q'^2+\lambda^2\phi'^2 +2\xi\left(\cos\phi-q'\sin\phi \right) \right. \notag\\
    &\left.\qquad+2\eta\left(\cos^2\phi-q'\sin2\phi\right) \right], \label{eq:varenergy}
\end{align}
where $\xi=\pi D/4\sqrt{AK}$ and $\eta=(\ln2)t_F\mu_0M_s^2/4\pi\sqrt{AK}$ are two dimensionless parameters characterizing the strengths of the DMI and the demagnetization energy relative to the DW energy $4\sqrt{AK}$. In typical ferromagnets, $\xi$ and $\eta$ are small so we only keep linear order terms for these parameters in Eq.~\eqref{eq:varenergy}.

Minimizing the free energy calls for two Euler-Lagrange equations. The first one $\delta_q E=0$ yields $q'-\xi\sin\phi-\eta\sin2\phi=C$ where the constant $C$ can be determined by the boundary conditions. At $y\rightarrow\pm\infty$, we have $q'\rightarrow0$ and $\phi\rightarrow-\arccos(\xi/2\eta)$, thus $C=0$. Including the other equation $\delta_\phi E=0$, we arrive at two coupled nonlinear differential equations
\begin{align}
    -\lambda^2\phi''&=\xi(\sin\phi+q'\cos\phi)+\eta(\sin2\phi+2q'\cos2\phi), \label{eq:phiprime}\\
    q'&=\xi\sin\phi+\eta\sin2\phi. \label{eq:qprime}
\end{align}
Since Eq.~\eqref{eq:varenergy} is accurate to linear order in $\xi$ and $\eta$, we ignore quadratic terms of $\xi$ and $\eta$ in Eq.~\eqref{eq:qprime} and Eq.~\eqref{eq:phiprime}, by which $\phi$ effectively decouples from $q$. Then Eq.~\eqref{eq:qprime} reduces to a double Sine-Gordon equation that, despite high non-linearity, can be solved analytically. Defining $\beta=\xi/2\eta$ as the relative strength of DMI with respect to the demagnetization energy, we obtain our central results:
\begin{align}
    \frac{\phi}2=
    \begin{cases}
        \pm\arctan\left[\sqrt{\frac{1+\beta}{1-\beta}}\tanh\left(\frac12\sqrt{1-\beta^2}y/\lambda_s\right)\right] \mbox{ if }\beta< 1 \\[12pt]
        \pm\arctan\left[\sqrt{\frac{\beta}{\beta-1}}\sinh\left(\sqrt{\beta-1}y/\lambda_s\right)\right] \mbox{ if }\beta\ge1
    \end{cases}
    \label{eq:DSGphi}
\end{align}
and
\begin{align}
    \frac{q}{\lambda_s}=\begin{cases}
     2\eta(1+\beta) \frac{\cosh\left(\sqrt{1-\beta^2}y/\lambda_s\right)-1}{\cosh\left(\sqrt{1-\beta^2}y/\lambda_s\right)-\beta} \quad \mbox{if }\beta<1 \\[12pt]
     4\eta\sqrt{\beta}\left[1+\frac{(\beta-1)\cosh\left(\sqrt{\beta-1}y/\lambda_s\right)}{1-\beta\cosh^2\left(\sqrt{\beta-1}y/\lambda_s\right)}\right] \mbox{ if }\beta\ge1
    \end{cases}
    \label{eq:DSGq}
\end{align}
where $\lambda_s=\lambda/\sqrt{2\eta}$ is an exchange length along $y$ at the critical point $\beta=1$ and the $+$ ($-$) sign represents the solution with positive (negative) topological charge $Q$. The critical condition $\beta=1$ is where the N\'{e}el wall is formed at $y\rightarrow\pm\infty$ and a DW skyrmion with $Q=\pm1$ is formed at the center. At this value, $\phi/2\rightarrow\pm\arctan y/\lambda_s$ and $q/\lambda_s\rightarrow\frac{4\eta(y/\lambda_s)^2}{1+(y/\lambda_s)^2}$. For $\beta<1$, only a partial DW Skrymion with $1/2 \le |Q|<1$ exists and $\phi(\pm\infty)=\pm[\pi-\arccos(\beta)]$. For $\beta=0$, Eqs.~\eqref{eq:DSGphi} and~\eqref{eq:DSGq} reduce to a VBL profile~\cite{Malozemoff1972,Slonczewski1974}. When converted into original units, the critical condition becomes
\begin{align}
    D_c=\frac{2\ln2}{\pi^2}t_F\mu_0M_s^2,
\end{align}
which sets a minimum DMI strength to form a full DW Skyrmion. To characterize the impact of DMI on the DW Skyrmion energy, $\Delta E$ (i.e. the energy cost of creating a DW Skyrmion inside a DW), we normalize by the VBL energy, $\Delta E_{\rm VBL}$.~\cite{Thiaville1995,Malozemoff1972,Slonczewski1974} This leads to a rather simple form of the scaled DW Skyrmion energy. 
\begin{align}
    \frac{\Delta E}{\Delta E_{\rm VBL}}=
    \begin{cases}
    \sqrt{1-\beta^2}+2\beta\arctan\sqrt{\frac{1+\beta}{1-\beta}}\quad\mbox{ if }\beta<1\\[12pt]
    2\sqrt{\beta-1}+2\beta\,\rm arccsc\sqrt{\beta}\quad\mbox{ if }\beta\ge1
    \end{cases}
\end{align}
which is plotted in Fig.~\ref{fig:DW-width-vs-D}b along with the corresponding micromagnetic calculations to be discussed below.

\newcommand{\sfwidth}{0.092\textwidth}

\begin{figure}[ht]
  \includegraphics[width=0.5\textwidth]{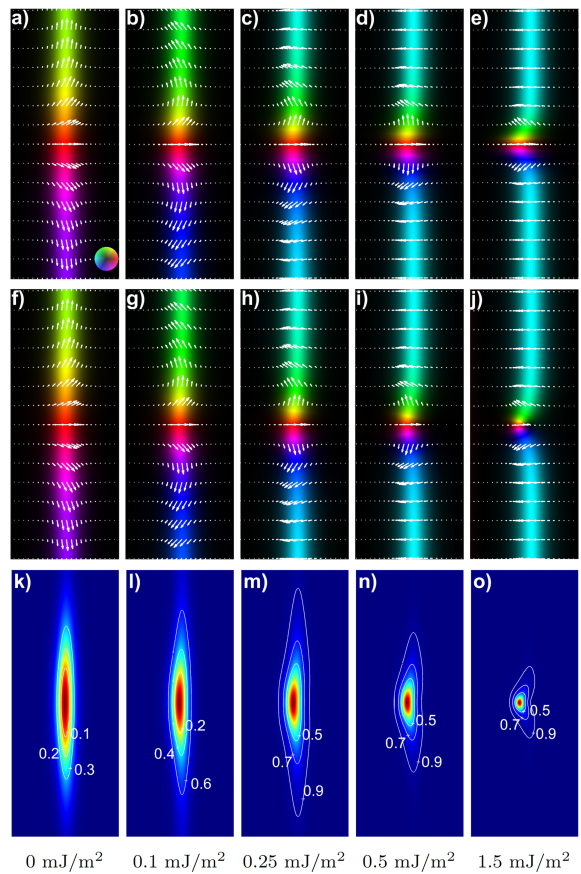}
  \caption{(a)-(e) Analytical solutions Eq.~\eqref{eq:DSGphi} and~\eqref{eq:DSGq}. (f)-(j) full micromagnetic solutions.  Arrows indicate in-plane component of the magnetization. (k)-(o) normalized topological charge densities - here the contour lines enclose the topological charge indicated. The DMI strength is indicated at the bottom of each column. The areas shown are 100 nm $\times$ 250 nm.}
  \label{fig:profile}
\end{figure}

\begin{figure}[ht]
  \centering
  \includegraphics[width=0.5\textwidth]{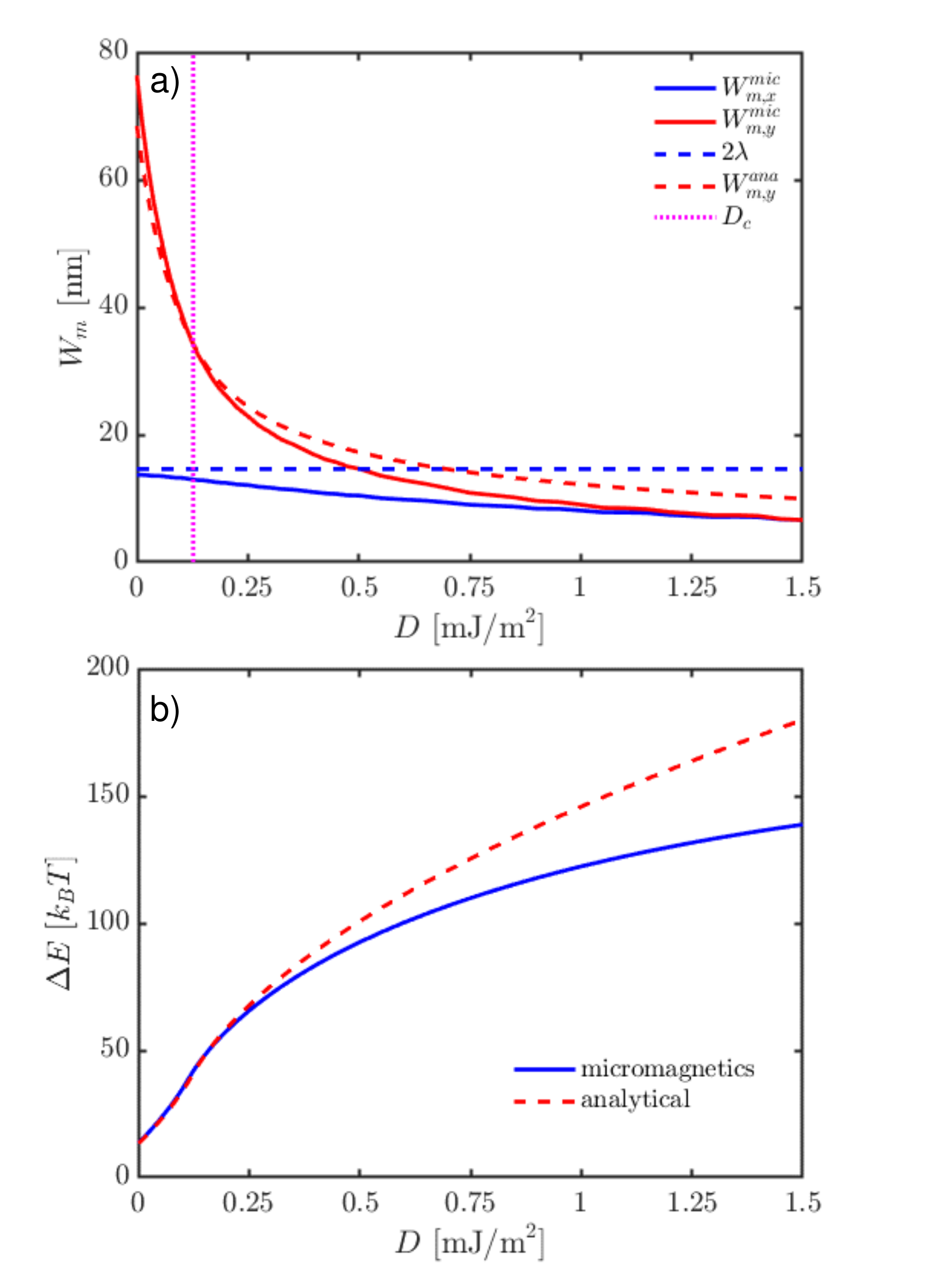}
  \caption{a) DW Skyrmion width $W_{m,x}$ (blue) across and $W_{m,y}$ (red) along the domain wall as a function of DMI strength for micromagnetic (solid lines) and analytical (dashed lines) calculations. The critical DMI strength $D_c$ is shown as a dotted magenta line. b) Corresponding DW Skyrmion energy vs DMI. Parameters used were $t_F=2$ nm, $K=3\times10^5$ J/m$^3$, $A=1.6\times10^{-11}$ J/m, and $M_s=600$ kA/m }
  \label{fig:DW-width-vs-D}
\end{figure}

\textit{Micromagnetic calculations.}---We used our MATLAB based finite differences code $\text{M}^3$~\cite{M3,Mohammadi2018}. The code implements the Dzyaloshinskii-Moriya interaction for thin films and the corresponding boundary conditions~\cite{Rohart2013} together with the exchange interaction for micromagnetics~\cite{Donahue2004}. As can be seen in Fig.~\ref{fig:profile} the magnetization profile of the analytic solutions agrees well with the full micromagnetic results, including the notch-like deformation near the center which ascribes to an increasing $D$. Parameters used in these calculations are as follows: $t_F=2$ nm, $K=3\times10^5$ J/m$^3$, $A=1.6\times10^{-11}$ J/m, and $M_s=600$ kA/m, which are comparable to values reported for Co/Ni multi-layers in \cite{Lau2018,Pellegren2017}.  The total volume simulated was 128 nm x 512 nm x 2 nm and the cell size was 0.5 nm x 0.5 nm x 2 nm.  In regard to future applications, the size of a Skyrmion plays an important role.  A conventional Skyrmion, i.e., a N\'eel or Bloch type Skyrmion, consists of an inner and outer domain as well as a DW separating them. The Skyrmion size is often given by its radius which is defined by the inner area bounded by the contour for which the out of plane magnetization vanishes, thereby neglecting the wall width~\cite{Wang2018}. Because the DW Skyrmion is confined within a distorted DW (Figure~\ref{fig:profile}), the conventional definition of a single Skyrmion radius is not applicable. However, one can use the DW width $W_{m,x}$ at the Skyrmion center and the width of the DW substructure $W_{m,y}$ along the wall to obtain an estimate of the size of the DW Skyrmion (for details, see supplemental materials).\footnote{See Supplemental Material at [URL will be inserted by publisher] for a definition of the DW Skyrmion widths, which includes refs \cite{Barzilai1988,Cauchy1847,Dai2003,Hubert1998}} As shown in Fig.~\ref{fig:DW-width-vs-D} both quantities decrease with increasing DMI. For the analytical solution, $\lambda$ appearing in Eq.~\eqref{eq:theta} and~\eqref{eq:phi} is assumed to be independent of $D$. It is simply given by $\lambda=\sqrt{A/K}$, which provides an upper bound for the width $W_{m,x}^{mic}$ from the micromagnetic simulations. As shown in Fig.~\ref{fig:DW-width-vs-D}, the analytical value $W_{m,y}^{ana}$ provides a good approximation for the width $W_{m,y}^{mic}$ determined from micromagnetic calculations. However, these two quantities do not capture the unique shape of DW Skyrmions. We therefore propose an alternative way to define the Skyrmion size and shape using the topological charge density~\cite{Heinze2011}:
\begin{align}
 \varrho_{\text{top}}=\frac{1}{4\pi}\bm{m}\cdot \left(\partial_x\bm{m}\times\partial_y\bm{m}\right).
\end{align}
The size of an arbitrary Skyrmion can now be defined as the area enclosing a certain percentage of the topological charge. Normalized plots of topological charge density are shown in Fig.~\ref{fig:profile}. The core of the DW Skyrmion defined by $W_{m,x}$ and $W_{m,y}$ contains about $20\%$ of the topological charge of the DW Skyrmion.  

Figure \ref{fig:DW-width-vs-D}b shows the DW Skyrmion energy vs D in units of $k_bT$ for $T=300K$ for analytical and micromagnetic calculations, which have near perfect agreement in the low D regime.  For larger D, the fixed $\lambda$ approximation becomes less valid causing the analytical solution to deviate from the micromagnetic one.

\begin{figure}
  \centering
  \includegraphics[width=0.48\textwidth]{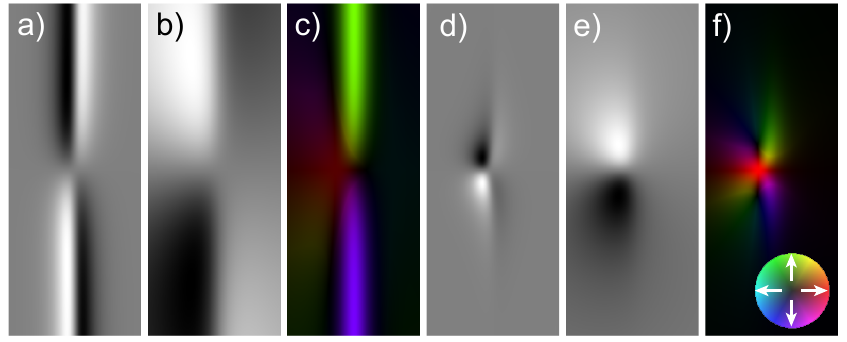}
  \caption{a \& d) Fresnel mode LTEM, b \& e) phase map, and c \& f) in-plane magnetic induction map of a a - c) vertical Bloch line (D = 0 mJ/m$^2$) and d - f) DW Skyrmion (D = 0.5 mJ/m$^2$) calculated from the micromagnetic output of figures \ref{fig:profile}f and \ref{fig:profile}i.}
  \label{fig:VBLLTEM}
\end{figure}

\textit{Lorentz TEM simulations.}---To support future experimental imaging of DW Skyrmions, we employ Fresnel mode Lorentz TEM calculations on the micromagnetic output of Fig.~\ref{fig:profile}. Fresnel mode Lorentz TEM is an out-of-focus imaging technique in which a through-focus series of bright field images is recorded; details regarding the simulation of relevant image contrast can be found in~\cite{degraef2000d}. Numerical profiles of an isolated VBL (D = 0 mJ/m$^2$) and an isolated DW Skyrmion (D = 0.5 mJ/m$^2$) are illustrated in Fig.~\ref{fig:profile}f and~i, respectively, which are used in the calculations of Fig.~\ref{fig:VBLLTEM}. In the absence of DMI, Bloch walls are present which display a sharp magnetic contrast that reverses at the location of the VBL in Fresnel mode images (figure~\ref{fig:VBLLTEM}a). In the presence of DMI, N\'eel walls become the preferred configuration and do not display any magnetic contrast in Fresnel mode images without sample tilt. However, strong magnetic contrast is still observed at the location of the DW Skyrmion in Fig.~\ref{fig:VBLLTEM}d. This dipole-like contrast originates from the Bloch-like portions of the DW across the DW Skyrmion. Thus, DW Skyrmions would be the only contributor to magnetic contrast in systems that exhibit DMI when examined with Lorentz TEM in the absence of sample tilt. 

As experimental Fresnel-mode Lorentz TEM images do not offer explicit directional information regarding the magnetic induction, phase reconstruction is typically employed using the Transport of Intensity Equation (TIE) to calculate the integrated in-plane magnetic induction~\cite{Paganin1998,Beleggia2003}. The resultant phase map for D=0 displays contrast along the domain wall which reverses at the location of the VBL similar to that observed in the Fresnel mode image. The color map shows the direction of in-plane induction matching those in the output of the micromagnetic simulation with a discontinuity at the location of the VBL.  The magnetic induction takes on a distinct braid-like appearance centered around the DW Skyrmion with no signal from the surrounding DW. This signature takes on a larger footprint than that of magnetic contrast in the calculated Fresnel mode image which may assist in locating DW Skyrmions in experimental images. 

\begin{figure}
  \centering
  \includegraphics[width=0.5\textwidth]{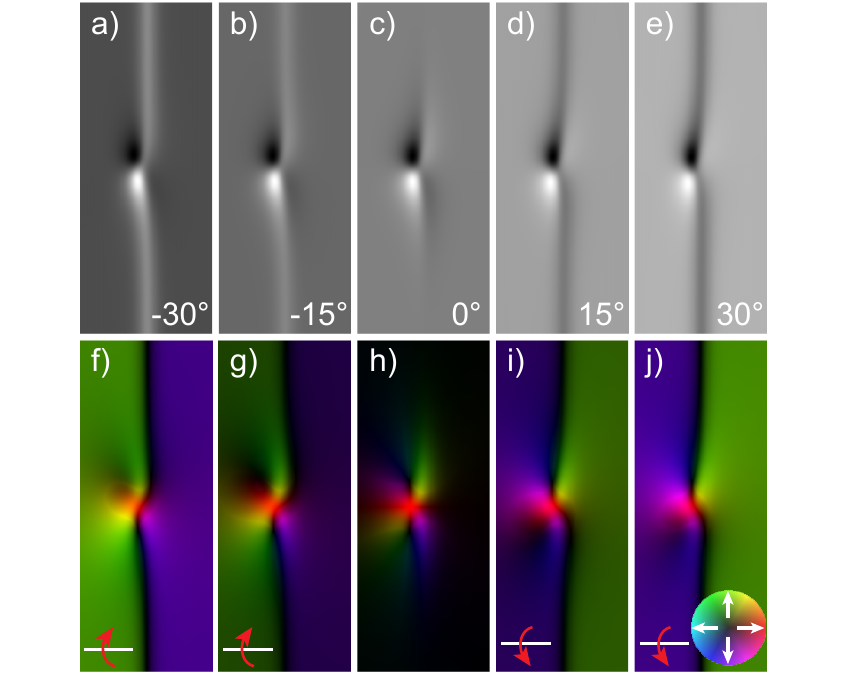}
  \caption{a - e) Fresnel mode LTEM images and f - j) corresponding in-plane magnetic induction maps of an isolated DW Skyrmion (D = 0.5 mJ/m$^2$) at varied states of tilt calculated from micromagentic outputs illustrated in figure \ref{fig:profile}i.}
  \label{fig:TILTLTEM}
\end{figure}

As mentioned previously, N\'eel walls do not display magnetic contrast in the absence of a sample tilt in Fresnel mode imaging. When a tilt is applied to the sample, an in-plane component emerges from the perpendicular induction of neighboring domains giving rise to contrast at a N\'eel wall. This too is observed in our calculated Fresnel mode images (Figure~\ref{fig:TILTLTEM}a-e); as sample tilt increases, magnetic contrast becomes more apparent along the DW surrounding the DW Skyrmion. Additionally, the contrast from the DW Skyrmion itself remains strong with respect to the surrounding DW regardless of tilt direction, which will be useful for confirming the presence of a DW Skyrmion experimentally. The corresponding in-plane magnetic induction maps (Figure~\ref{fig:TILTLTEM}f-j) further support this notion as the braid-like feature from the DW Skyrmion remains visible even at larger tilts where a strong signal is observed around the DW.

In summary, we have introduced a new kind of topological magnetic excitation called a DW Skyrmion characterized by a $360^\circ$ transition of the internal magnetization within a Dzyaloshinski DW and defined by a topological charge of $\pm1$.  The DW Skyrmion analysis presented here builds off prior work on VBLs in much the same way the recent surge in Skyrmion research is rooted in decades of research on magnetic bubble memory.  The static properties were calculated both analytically and micromagnetically with excellent agreement on the resulting size, energy, and profile.  Although open questions remain about their thermal stability and dynamic properties, DW Skyrmions provide an alternative strategy for leveraging topological protection in magnetic systems with a strong interfacial DMI.  The reconfigurable nature of the DWs that host these excitations could open the door to new kinds of memory and computing schemes based on topological charge. To this end, we have proposed an experimental methodology to unequivocally image DW Skyrmions using Fresnel mode Lorentz TEM to support future work in this area.

\section{Acknowledgements}
This work is financially  supported by the Defense Advanced Research Project Agency (DARPA) program on Topological Excitations in Electronics (TEE) under grant number D18AP00011. C.M. and A.S. would also like to acknowledge support by NSF-CAREER grant \#1452670.

\bibliography{citations}

\end{document}